\documentclass[aps,PRB,twocolumn,superscriptaddress,preprintnumbers]{revtex4-2}
\usepackage[T1]{fontenc}
\usepackage{times}
\usepackage{graphicx,color}
\usepackage{amsfonts,amsmath,amssymb,amsbsy}
\usepackage[colorlinks=true, linkcolor=blue, citecolor=blue, urlcolor=blue]{hyperref}
\usepackage{float}

\begin{document}

\title{Quantum geometry and elliptic optical dichroism in $p$-wave magnets}
\author{Motohiko Ezawa}
\affiliation{Department of Applied Physics, The University of Tokyo, 7-3-1 Hongo, Tokyo
113-8656, Japan}

\begin{abstract}
The quantum geometric tensor is composed of the Berry curvature and the
quantum metric, which is observable by means of optical absorption of
elliptically polarized light. Especially, the quantum geometric tensor at
the zero-momentum is observable by the optical absorption at the optical
band edge. In this context, we study optical absorption of a $p$-wave magnet
under irradiation of elliptically polarized light. The $p$-wave magnet has a
band splitting along one axis, which we choose the $x$\ axis. We obtain
analytic formulae for the optical conductivity up to the second order in the
magnitude of the N\'{e}el vector. In particular, the optical conductivity is
exactly obtained when the N\'{e}el is along the $x$, $y$ and $z$ axis. It
shows strong ellipticity a dependence of the light polarization, which is an
elliptic dichroism. Especially, there is a perfect elliptic optical
dichroism when the N\'{e}el vector is along the $y$ axis. It is possible to
determine the N\'{e}el vector by measuring the ellipticity of the perfect
elliptic dichroism.
\end{abstract}

\date{\today }
\maketitle

\section{Introduction}

Optical absorption experiments provide us with a powerful tool to observe
material properties. For example, the band gap is determined by the edge of
the optical absorption bands. Especially, circular optical dichroism of the
gapped Dirac system can determine the chirality of the Dirac cones\cite%
{Yao08,Xiao,EzawaOpt,Li}. For example, there is no optical absorption at the
band edge for the left-polarized light but there is a finite optical
absorption for the right-polarized light. In a similar way, elliptic optical
dichroism was studied in the anisotropic Dirac system\cite{TCIOpt}.

Quantum geometry attracts renewed attention in condensed-matter physics.
Recent studies demonstrate that they appear in optical conductivity\cite%
{Ahn,Holder,Bhalla,Ahn2,Souza,WChen2022,Onishi,Sousa,Ghosh,WChen2024,EzawaQG,Oh}
and electric nonlinear conductivity\cite%
{MaNature,CWang,KamalDas,AGao,NWang,Kaplan,Sala,YFang,EzawaAlter}. The
quantum geometry is characterized by the quantum geometric tensor, whose
imaginary part is the Berry curvature and real part is the quantum metric.
The Berry curvature is observable by the anomalous Hall effect. On the other
hand, it is still a nontrivial problem to observe the quantum metric. It is
possible to observe the quantum metric by the optical conductivity under
linearly polarized light\cite{Souza,WChen2022,Onishi,Sousa,Ghosh,EzawaQG,Oh},

The band structure of the $p$-wave magnet has the $p$-wave symmetry, whose
net magnetization is zero\cite{pwave}. The electron coupled with the $p$%
-wave magnet feels a $p$-wave symmetric field. Recently, an experiment on
the $p$-wave magnet Gd$_{3}$Ru$_{4}$Al$_{12}$\ was reported\cite{Yamada}.
Especially, the $p$-wave magnet at the interface is studied\cite%
{EzawaPwave,EzawaPNeel,PEdel}, where the Rashba interaction is present. In
general, it is difficult to read out the N\'{e}el vector of the $p$-wave
magnet due to the absence of the net magnetization. It was pointed out that
this is possible by measuring linear and nonlinear conductivities. It is
benefitable if the N\'{e}el vector of the $p$-wave magnet can be read out
solely by optical absorption.

In this paper, we show that the quantum metric and the Berry curvature is
observable by the optical absorption under elliptically polarized light.
Especially, the quantum metric and the Berry curvature at the zero momentum
is observed by the optical absorption at the optical band edge. We
explicitly study optical absorption in a $p$-wave magnet coupled with the
Rashba interaction. The $p$-wave magnet has a band splitting along one axis,
which we choose the $x$\ axis. We analytically calculate the optical
conductivity under irradiation of elliptically polarized light. Especially,
the optical conductivity is exactly obtained when the N\'{e}el is along the $%
x$, $y$ and $z$ axis. It shows strong ellipticity dependence of the light
polarization, which is an elliptic dichroism. When the N\'{e}el vector is
along the $y$ axis, the perfect elliptic dichroism occurs, where there is no
optical absorption at the optical band edge for a certain right-polarized
light but there is a finite optical absorption for a corresponding
left-polarized light.

This paper is composed as follows. In Sec.II, we introduce a model for a $p$%
-wave magnet and show the energy spectrum. In Sec.III, we review the optical
absorption under elliptically polarized light. In Sec.IV, we relate the
optical absorption under the elliptically polarized light and quantum
geometric properties including the Berry curvature and the quantum metric.
In Sec.V, we calculate the Berry curvature and the quantum metric for the $p$%
-wave magnet. In Sec.VI, we explicitly calculate the optical absorpition
under elliptically polarized light for the $p$-wave magnets. We find a
perfect elliptic dichroism occurs when the N\'{e}el is along the $y$
direction. Sec.VII is devoted to discussions on possible applications.

\section{$p$-wave magnet}

We consider a $p$-wave magnet on a substrate. The Rashba interaction is
introduced by placing a sample on the substrate\cite%
{SmejRev,SmejX2,Zu2023,Sun,EzawaAlter,Diniz,Rao,Amund,EzawaPNeel}. The
Hamiltonian is typically given by\cite{EzawaPNeel,PEdel,Yamada},%
\begin{equation}
H\left( \mathbf{k}\right) =\frac{\hbar ^{2}\left( k_{x}^{2}+k_{y}^{2}\right) 
}{2m}\sigma _{0}+\lambda \left( k_{x}\sigma _{y}-k_{y}\sigma _{x}\right)
+k_{x}\mathbf{J}\cdot \mathbf{\sigma }+B\sigma _{z},  \label{pHamil}
\end{equation}%
where $m$ is the effective mass of free electrons, $\lambda $ is the
magnitude of the Rashba interaction, and 
\begin{equation}
\mathbf{J}=J\left( \sin \Theta \cos \Phi ,\sin \Theta \sin \Phi ,\cos \Theta
\right)
\end{equation}%
is the $p$-wave N\'{e}el vector with $J$ its magnitude\cite%
{pwave,Okumura,EzawaPwave,Brek,EzawaPNeel,GIAlter}. We assume that $%
J<\left\vert \lambda \right\vert $ so that the Rashba interaction forms a
Dirac cone at the Dirac point $k_{x}=k_{y}=0$. Additionally, we have
introduced the $B\sigma _{z}$ term to introduce the band gap at the Dirac
point, which is introduced by applying an external magnetic field or
attaching a ferromagnet.

\begin{figure}[t]
\centerline{\includegraphics[width=0.49\textwidth]{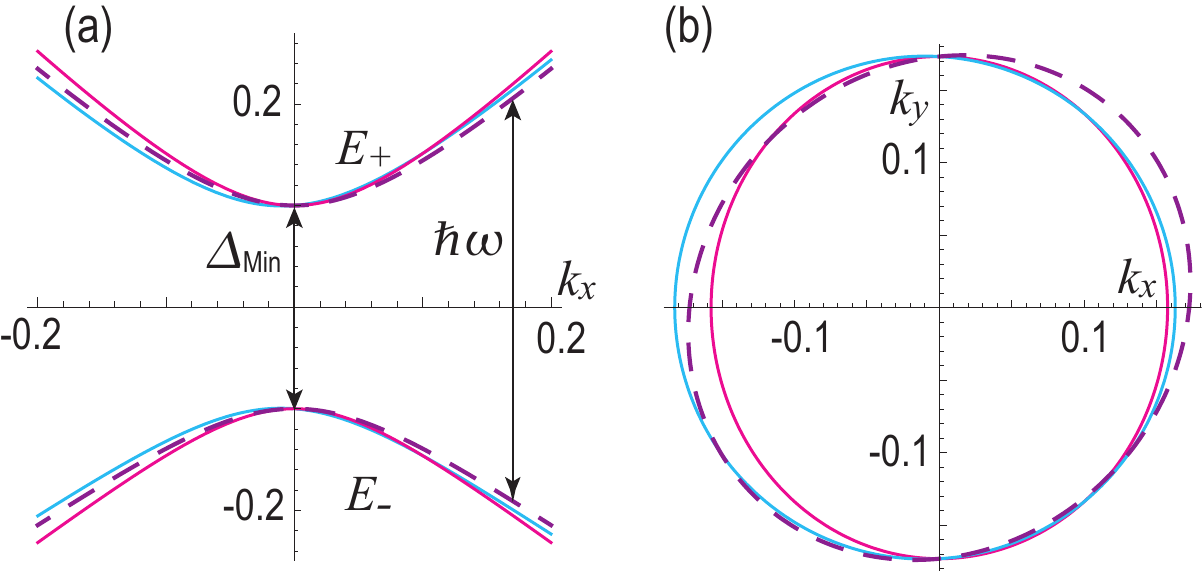}}
\caption{(a) Band structure along the $x$ axis. The horizontal axis is $%
-0.2k_{0}<k<0.2k_{0}$. The vertical axis is the energy in units of $\protect%
\varepsilon _{0}$. (b) Fermi surface at $\protect\omega =0.4\protect%
\varepsilon _{0}/\hbar $. The N\'{e}el vector is along the $y$ axis ($\Theta
=\protect\pi /2$, $\Phi =\protect\pi /2$) for magenta curves, along the $x$
axis ($\Theta =\protect\pi /2$, $\Phi =0$) for dashed purple curves, and
along the $z$ axis ($\Theta =0$) for cyan curves. We have set $J=0.1\protect%
\varepsilon _{0}/k_{0}$, $\protect\lambda =\protect\varepsilon _{0}/k_{0}$, $%
B=0.1\protect\varepsilon _{0}$ and $m=\hbar ^{2}k_{0}^{2}/2$, where $\protect%
\varepsilon _{0}$ is a unit of the energy and $k_{0}$ is a unit of the
momentum.}
\label{FigPBand}
\end{figure}

The energy of the Hamiltonian (\ref{pHamil}) is given by%
\begin{equation}
E_{\pm }\left( \mathbf{k}\right) =\frac{\hbar ^{2}\left(
k_{x}^{2}+k_{y}^{2}\right) }{2m}\pm \Delta \left( \mathbf{k}\right) ,
\end{equation}%
with%
\begin{equation}
\Delta \left( \mathbf{k}\right) =\sqrt{\left( J_{x}k_{x}-\lambda
k_{y}\right) ^{2}+\left( J_{y}k_{x}+\lambda k_{x}\right) ^{2}+\left(
J_{z}k_{x}+B\right) ^{2}}.
\end{equation}%
The band structure along the $x$ axis is shown in Fig.\ref{FigPBand}(a).

The free electron term $\hbar ^{2}k^{2}/\left( 2m\right) $ does not play an
important role compared with the interband transition $2\Delta \left( 
\mathbf{k}\right) $ in optical absorption, as is understood from the formula
(\ref{absorp}) which we derive later.

\section{Optical absorption}

We study optical inter-band transitions from the state $|\psi _{-}(\mathbf{k}%
)\rangle $ in the valence band to the state $|\psi _{+}(\mathbf{k})\rangle $
in the conduction band. We apply a beam of elliptical polarized light
perpendicular onto the sample, where the corresponding electromagnetic
potential is given by $\mathbf{A}(t)=(A_{x}\sin \omega t,A_{y}\cos \omega t)$%
. The electromagnetic potential is introduced into the Hamiltonian by way of
the minimal substitution, that is, by replacing the momentum $k_{j}$ with
the covariant momentum $P_{j}\equiv k_{j}+eA_{j}$.

We start with the optical matrix element between the initial and final
states in the photo-emission process given by\cite%
{Yao08,Xiao,EzawaOpt,Li,TCIOpt} 
\begin{equation}
P_{i}(\mathbf{k})\equiv \hbar \left\langle \psi _{+}(\mathbf{k})\right\vert
v_{\mu }\left\vert \psi _{-}(\mathbf{k})\right\rangle ,  \label{EqP}
\end{equation}%
with the velocity%
\begin{equation}
v_{\mu }=\frac{\partial H\left( \mathbf{k}\right) }{\hbar \partial k_{\mu }}.
\end{equation}%
The optical matrix element of the elliptic polarization is given by%
\begin{equation}
P_{\vartheta }(\mathbf{k})=P_{x}(\mathbf{k})\cos \vartheta +iP_{y}(\mathbf{k}%
)\sin \vartheta ,
\end{equation}%
where $\vartheta $ is the ellipticity of the injected beam, with $%
0<\vartheta <\pi $ for the right polarization and $-\pi <\vartheta <0$ for
the left polarization. $P_{\pi /4}(\mathbf{k})$ corresponds to the right
circularly polarized right and $P_{-\pi /4}(\mathbf{k})$ corresponds to the
left circularly polarized right,

Optical absorption is given by the optical conductivity%
\begin{align}
\sigma \left( \omega ;\vartheta \right) =& \frac{\sigma _{0}}{\omega ^{2}}%
\int d\mathbf{k}\left[ f_{-}(\mathbf{k})-f_{+}(\mathbf{k})\right] \left\vert
P_{\vartheta }(\mathbf{k})\right\vert ^{2}  \notag \\
& \times \delta \left[ E_{+}(\mathbf{k})-E_{-}(\mathbf{k})-\hbar \omega %
\right] ,  \label{absorp}
\end{align}%
where $f_{\pm }(\mathbf{k})=1/\left( \exp \left[ \left( E_{\pm }(\mathbf{k}%
)-\mu \right) /\left( k_{\text{B}}T\right) +1\right] \right) $ is the Fermi
distribution function and we have set%
\begin{equation}
\sigma _{0}\equiv \frac{\pi e^{2}}{\left( 2\pi \right) ^{2}\varepsilon _{0}}.
\end{equation}%
In the following, we assume that the temperature is absolutely zero.

\begin{figure}[t]
\centerline{\includegraphics[width=0.49\textwidth]{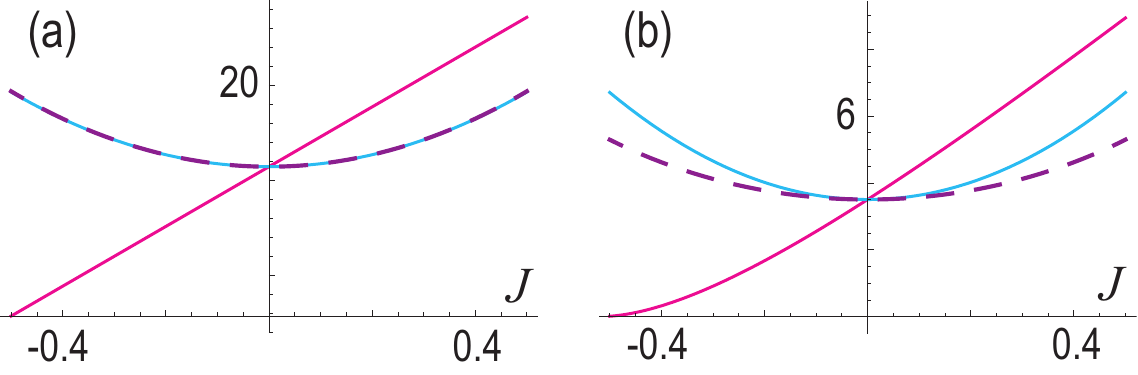}}
\caption{Optical conductivity at the optical band edge as a function of $J$
under linear polarized light for (a) $\protect\vartheta =0$ and (b) $\protect%
\vartheta =\protect\pi /2$. See the caption of Fig.1 for the colors. We have
set $\protect\lambda =\protect\varepsilon _{0}/k_{0}$, $B=0.1\protect%
\varepsilon _{0}$ and $m=\hbar ^{2}k_{0}^{2}/2$, where $\protect\varepsilon %
_{0}$ is a unit of the energy and $k_{0}$ is a unit of the momentum.}
\label{FigLinear}
\end{figure}

\section{Quantum geometry and elliptic dichroism}

The quantum geometric tensor is defined by\cite{Provost,Ma}%
\begin{equation}
\mathcal{F}_{mn}^{\mu \nu }\left( \mathbf{k}\right) =\left\langle \partial
_{k_{\mu }}\psi _{m}\left( \mathbf{k}\right) \right\vert 1-\mathcal{P}\left( 
\mathbf{k}\right) \left\vert \partial _{k_{\nu }}\psi _{n}\left( \mathbf{k}%
\right) \right\rangle ,
\end{equation}%
where 
\begin{equation}
\mathcal{P}\left( \mathbf{k}\right) \equiv \sum_{n\in \text{Occupied}%
}\left\vert u_{n}\left( \mathbf{k}\right) \right\rangle \left\langle
u_{n}\left( \mathbf{k}\right) \right\vert ,
\end{equation}%
is the projection operator to the occupied band. In the two band model, its
diagonal component is simply given by%
\begin{equation}
\mathcal{F}_{--}^{\mu \nu }\left( \mathbf{k}\right) =\left\langle \partial
_{k_{\mu }}\psi _{-}(\mathbf{k})\left\vert \psi _{+}(\mathbf{k}%
)\right\rangle \right. \left. \left\langle \psi _{+}(\mathbf{k})\right\vert
\partial _{k_{\nu }}\psi _{-}(\mathbf{k})\right\rangle .
\end{equation}%
The optical matrix element for the ellipcit polarized light is expanded as%
\begin{align}
& \left\vert P_{\vartheta }(\mathbf{k})\right\vert ^{2}  \notag \\
=& \left\vert P_{x}(\mathbf{k})\right\vert ^{2}\cos ^{2}\theta +\left\vert
P_{y}(\mathbf{k})\right\vert ^{2}\sin ^{2}\theta  \notag \\
& +i\left[ P_{x}^{\ast }(\mathbf{k})P_{y}(\mathbf{k})-P_{y}^{\ast }(\mathbf{k%
})P_{x}(\mathbf{k})\right] \sin \theta \cos \theta .
\end{align}%
By using the Hellmann-Feynman theorem%
\begin{align}
& \left\langle \psi _{m}\left( \mathbf{k}\right) \right\vert v_{\mu
}\left\vert \psi _{n}\left( \mathbf{k}\right) \right\rangle  \notag \\
& =\frac{1}{\hbar }\left( E_{n}\left( \mathbf{k}\right) -E_{m}\left( \mathbf{%
k}\right) \right) \left\langle \psi _{m}\left( \mathbf{k}\right) \right\vert
\partial _{k_{\mu }}\left\vert \psi _{n}\left( \mathbf{k}\right)
\right\rangle ,\quad
\end{align}%
for$\quad m\neq n$, we have%
\begin{align}
\left\vert P_{\mu }(\mathbf{k})\right\vert ^{2}=& \hbar ^{2}\left\langle
\psi _{-}(\mathbf{k})\right\vert v_{\mu }\left\vert \psi _{+}(\mathbf{k}%
)\right\rangle \left\langle \psi _{+}(\mathbf{k})\right\vert v_{\mu
}\left\vert \psi _{-}(\mathbf{k})\right\rangle  \notag \\
=& -\Delta ^{2}\left( \mathbf{k}\right) \left\langle \partial _{k_{\mu
}}\psi _{-}(\mathbf{k})\left\vert \psi _{+}(\mathbf{k})\right\rangle \right.
\left. \left\langle \psi _{+}(\mathbf{k})\right\vert \partial _{k_{\mu
}}\psi _{-}(\mathbf{k})\right\rangle  \notag \\
=& -\Delta ^{2}\left( \mathbf{k}\right) g_{\mu \mu }\left( \mathbf{k}\right)
,
\end{align}%
with $\mu =x,y$ and%
\begin{align}
& i\left[ P_{y}^{\ast }(\mathbf{k})P_{x}(\mathbf{k})-P_{x}^{\ast }(\mathbf{k}%
)P_{y}(\mathbf{k})\right]  \notag \\
& =i\Delta ^{2}\left( \mathbf{k}\right) [\left\langle \psi _{-}(\mathbf{k}%
)\right\vert v_{y}\left\vert \psi _{+}(\mathbf{k})\right\rangle \left\langle
\psi _{+}(\mathbf{k})\right\vert v_{x}\left\vert \psi _{-}(\mathbf{k}%
)\right\rangle  \notag \\
& \emph{\qquad \qquad }-\left\langle \psi _{-}(\mathbf{k})\right\vert
v_{x}\left\vert \psi _{+}(\mathbf{k})\right\rangle \left\langle \psi _{+}(%
\mathbf{k})\right\vert v_{y}\left\vert \psi _{-}(\mathbf{k})\right\rangle ] 
\notag \\
& =i\Delta ^{2}\left( \mathbf{k}\right) \left[ \mathcal{F}_{-+}^{xy}\left( 
\mathbf{k}\right) -\mathcal{F}_{+-}^{yx}\left( \mathbf{k}\right) \right]
=\Delta ^{2}\left( \mathbf{k}\right) \Omega _{xy}\left( \mathbf{k}\right) .
\end{align}%
Then, the optical conductivity is rewritten by using the quantum geometric
information as%
\begin{equation}
\sigma \left( \omega ;\vartheta \right) =\hbar ^{2}\sigma _{0}\int d\mathbf{k%
}f(\mathbf{k})G(\mathbf{k})\delta \left[ E_{+}(\mathbf{k})-E_{-}(\mathbf{k}%
)-\hbar \omega \right] ,  \label{OptG}
\end{equation}%
with%
\begin{equation}
G(\mathbf{k};\vartheta )\equiv g_{xx}(\mathbf{k})\cos ^{2}\vartheta +g_{yy}(%
\mathbf{k})\sin ^{2}\vartheta +\Omega _{xy}(\mathbf{k})\sin \vartheta \cos
\vartheta .
\end{equation}

\section{Quantum geometry in $p$-wave magnets}

In the two-band model whose Hamiltonian is given by the form%
\begin{equation}
H=h_{0}+\sum_{j=x,y,z}h_{j}\sigma _{j},
\end{equation}%
with $\sigma _{j}$ the Pauli matrix. The energy is spectrum is given by%
\begin{equation}
E\left( \mathbf{k}\right) =\sqrt{\sum_{j=0}^{N}h_{j}^{2}\left( \mathbf{k}%
\right) }.
\end{equation}%
The quantum metric for the two-band system is explicitly given by\cite%
{Matsuura,Gers,Onishi,WChen2024,EzawaQG,Oh}%
\begin{equation}
g_{\mu \nu }\left( \mathbf{k}\right) =\frac{1}{2}\left( \partial _{k_{\mu }}%
\mathbf{n}\right) \cdot \left( \partial _{k_{\nu }}\mathbf{n}\right) ,
\label{Gmn}
\end{equation}%
where $n_{j}\left( \mathbf{k}\right) =h_{j}\left( \mathbf{k}\right) /E\left( 
\mathbf{k}\right) \ $is the normalized Dirac vector with the energy. The
Berry curvature for the two-band system is explicitly given by\cite%
{Hsiang,Stic,Jiang}%
\begin{equation}
\Omega _{xy}=-\frac{1}{2}[\mathbf{n}\cdot (\partial _{x}\mathbf{n}\times
\partial _{y}\mathbf{n})].
\end{equation}%
The quantum metrics for the Hamiltonian (\ref{pHamil}) are given by 
\begin{align}
g_{xx}\left( \mathbf{k}\right) & =\frac{\left( \lambda
k_{y}J_{z}+BJ_{x}\right) ^{2}+\left( \lambda ^{2}k_{y}^{2}+B^{2}\right)
\left( \lambda +J_{y}\right) ^{2}}{2E^{4}\left( \mathbf{k}\right) },  \notag
\\
g_{yy}\left( \mathbf{k}\right) & =\lambda ^{2}\frac{\left(
B+J_{x}k_{x}\right) ^{2}+k_{x}^{2}\left( \lambda +J_{y}\right) ^{2}}{%
2E^{4}\left( \mathbf{k}\right) }.
\end{align}%
The Berry curvature for the Hamiltonian (\ref{pHamil}) is given by%
\begin{equation}
\Omega _{xy}\left( \mathbf{k}\right) =-\frac{B\lambda \left( \lambda
+J_{y}\right) }{2E^{3}\left( \mathbf{k}\right) }.
\end{equation}

\section{Optical conductivity in $p$-wave magnets}

The optical conductivity (\ref{absorp}) is rewritten as

\begin{equation}
\sigma \left( \omega ;\vartheta \right) =\sigma _{0}\int kdkd\phi \frac{G(%
\mathbf{k})}{2\left\vert \partial _{k}\Delta \mathbf{k}\right\vert }\delta
\left( k-k_{0}\left( \phi \right) \right) ,
\end{equation}%
where $k_{0}\left( \phi \right) $ is the solution of the resonant condition $%
2\Delta \left( k_{0}\left( \phi \right) \right) =\hbar \omega $ and $%
k_{x}=k\cos \phi $ and $k_{y}=k\sin \phi $. We show $k_{0}\left( \phi
\right) $ in Fig.\ref{FigPBand}(b).

We have%
\begin{align}
\frac{\partial H}{\partial k_{x}}& =\frac{\hbar ^{2}k_{x}}{m}+\lambda
k_{x}\sigma _{y}+J\mathbf{n}\cdot \mathbf{\sigma }, \\
\frac{\partial H}{\partial k_{y}}& =\frac{\hbar ^{2}k_{y}}{m}-\lambda \sigma
_{x}.
\end{align}%
In the optical absorption the energy difference of the conduction and
valence bands $\Delta \left( \mathbf{k}\right) \equiv E_{+}\left( \mathbf{k}%
\right) -E_{-}\left( \mathbf{k}\right) $ are important because the minimum
of $\Delta \left( \mathbf{k}\right) $ is 
\begin{equation}
\Delta _{\text{Min}}=\frac{2\left\vert B\left( \lambda +J_{y}\right)
\right\vert }{\sqrt{\left( \lambda +J_{y}\right) ^{2}+J_{z}^{2}}}
\end{equation}%
at 
\begin{align}
k_{x}^{\text{Min}}& =-\frac{BJ_{z}}{J_{z}^{2}+\left( \lambda +J_{y}\right)
^{2}},  \notag \\
k_{y}^{\text{Min}}& =-\frac{BJ_{z}J_{x}}{\lambda \left( J_{z}^{2}+\left(
\lambda +J_{y}\right) ^{2}\right) }.
\end{align}%
Especially, if $J_{z}=0$, the band edge is taken at the Dirac point ($%
k_{x}=k_{y}=0$), where the band gap is given by $2\Delta \left( 0\right)
=2\left\vert B\right\vert $.

\subsection{General case}

First we study the general case, where all of $J_{x}$, $J_{y}$ and $J_{z}$
are nonzero. The optical conductivity is given by 
\begin{align}
& \frac{\sigma \left( \omega ;\vartheta \right) }{\sigma _{0}}  \notag \\
=& \frac{\pi }{2\lambda ^{2}\hbar \omega ^{3}}\Big[\Big\{\Big(\lambda
^{2}+\lambda J_{y}+J_{x}^{2}-\frac{J_{z}^{2}}{2}\Big)(\hbar ^{2}\omega
^{2}+4B^{2})  \notag \\
& +4\pi B^{2}J_{z}^{2}\Big\}\cos ^{2}\vartheta  \notag \\
& +\Big\{\Big(\lambda ^{2}-\lambda J_{y}+J_{y}^{2}-\frac{J_{z}^{2}}{2}\Big)%
(\hbar ^{2}\omega ^{2}+4B^{2})  \notag \\
& -4\pi B^{2}J_{z}^{2}\Big\}\sin ^{2}\vartheta \Big]  \notag \\
& -\frac{2B\pi \left( 2\lambda ^{2}+J_{z}^{2}\right) }{\lambda ^{2}\omega
^{2}}\cos \vartheta \sin \vartheta
\end{align}%
up to the second order in $J$. In general, it is possible to determine $%
J_{x} $, $J_{y}$, $J_{z}$, $\lambda $ and $B$ by extracting them from the
band gap $2\Delta _{\text{Min}}$ and the $\vartheta $ dependence of the
optical conductivity at the optical band edge $\sigma \left( 2\Delta _{\text{%
Min}};\vartheta \right) $. The optical conductivity is quadratic in $J_{x}$\
and $J_{z}$\ but not in $J_{y}$\ as shown in Fig.\ref{FigLinear}(a) and (b).
In the following, we investigate the case where the N\'{e}el vector is along
the $y$, $x$ and $z$ axis, respectively.

\begin{figure}[t]
\centerline{\includegraphics[width=0.49\textwidth]{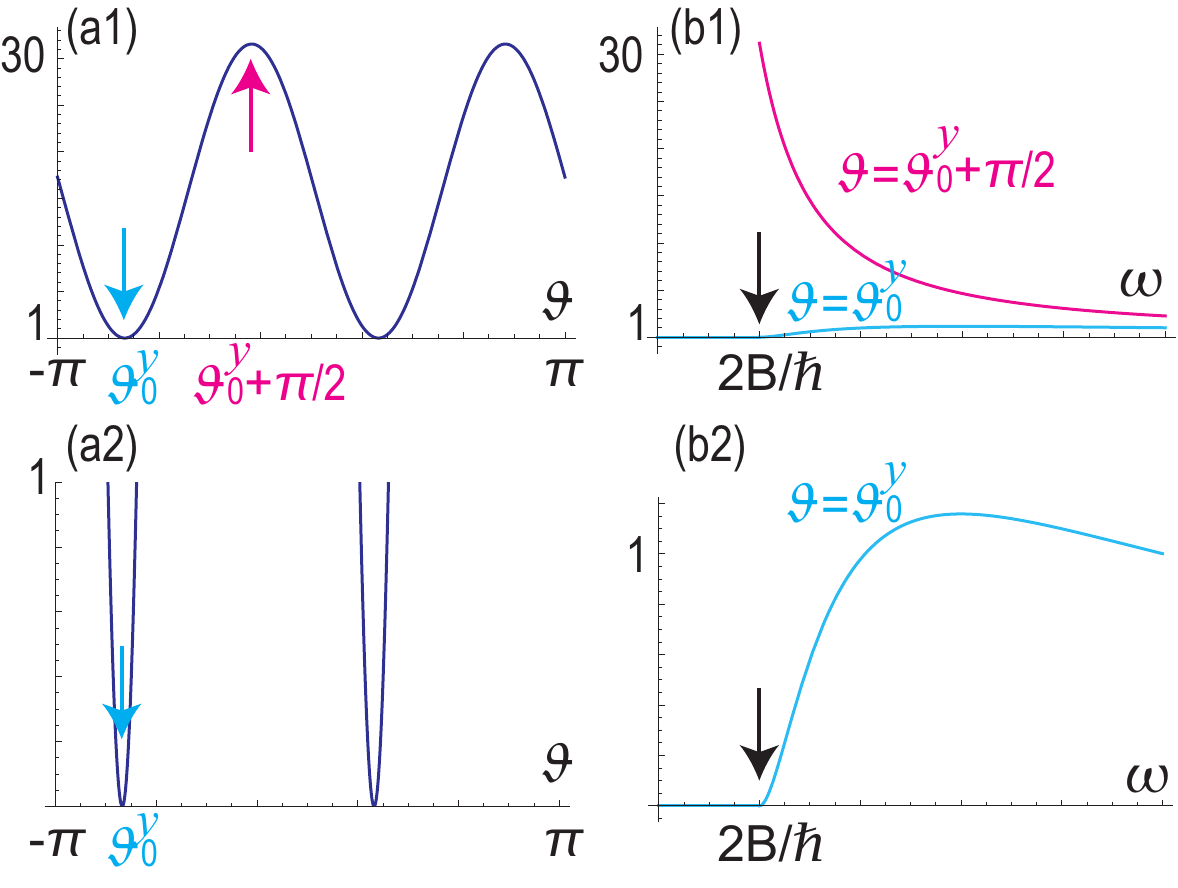}}
\caption{Elliptic dichroism, where the N\'{e}el vector is taken along the $y$
axis. (a1) and (a2) Optical conductivity as a function of $\protect\vartheta 
$. The horizontal axis is $\protect\vartheta $. The vertical axis is $%
\protect\sigma \left( \protect\omega \right) $ in units of $\protect\sigma %
_{0}$. (b1) and (b2) Optical conductivity as a function of $\protect\omega $%
. Magenta curves represents the $\protect\sigma \left( \protect\omega %
\right) $ with $\protect\vartheta =\protect\vartheta _{0}^{y}+\protect\pi /2$%
, while cyan curves represents $\protect\sigma \left( \protect\omega \right) 
$ with $\protect\vartheta =\protect\vartheta _{0}^{y}$. (a2) and (b2) are
the enlarged figures of (a1) and (b1) showing $\protect\sigma \left(
2\left\vert B\right\vert /\hbar ;\protect\vartheta _{0}^{y}\right) =0$. The
horizontal axis is $\hbar \protect\omega $. We have set $J=0.1\protect%
\varepsilon _{0}/k_{0}$, $\protect\lambda =\protect\varepsilon _{0}/k_{0}$
and $B=0.1\protect\varepsilon _{0}$, where $\protect\varepsilon _{0}$ is a
unit of the energy and $k_{0}$ is a unit of the momentum.}
\label{FigEllipticPy}
\end{figure}

\subsection{In-plane case\textbf{\ }$\mathbf{J}=(J_{x},J_{y},0)$}

We study the case where the N\'{e}el vector is along the in-plane direction.
In this case, the optical conductivity at the optical band edge is solely
determined by the contribution $\mathbf{k}=\mathbf{0}$ as%
\begin{equation}
\frac{\sigma \left( 2\left\vert B\right\vert ;\vartheta \right) }{\sigma _{0}%
}=\int_{0}^{2\pi }d\phi \frac{G(\mathbf{0};\vartheta )}{2\left\vert \partial
_{k}\Delta \mathbf{k}\right\vert _{\mathbf{k=0}}}=\frac{G(\mathbf{0}%
;\vartheta )}{2\left\vert \lambda \left( \lambda +J_{y}\right) \right\vert },
\end{equation}%
where we have used 
\begin{equation}
\partial _{k}\Delta \mathbf{k}=\frac{k}{E\left( \mathbf{k}\right) }[\left(
J_{x}\cos \phi -\lambda \sin \phi \right) ^{2}+\left( \lambda +J_{y}\right)
^{2}\cos ^{2}\phi ].
\end{equation}%
The quantum metrics at the Dirac point are%
\begin{equation}
g_{xx}\left( \mathbf{0}\right) =\frac{J_{x}^{2}+\left( \lambda +J_{y}\right)
^{2}}{32B^{2}},\quad g_{yy}\left( \mathbf{0}\right) =\frac{\lambda ^{2}}{%
32B^{2}}.
\end{equation}%
The Berry curvature at the Dirac point is%
\begin{equation}
\Omega _{xy}\left( \mathbf{0}\right) =-\frac{\lambda \left( \lambda
+J_{y}\right) }{16B^{2}}.
\end{equation}%
Then, we have%
\begin{align}
& G(\mathbf{0};\vartheta )  \notag \\
& =\frac{J_{x}^{2}}{32B^{2}}\cos ^{2}\vartheta +\frac{\left( \lambda
+J_{y}\right) ^{2}}{32B^{2}}\cos ^{2}\vartheta +\frac{\lambda ^{2}}{32B^{2}}%
\sin ^{2}\vartheta  \notag \\
& -\frac{\lambda \left( \lambda +J_{y}\right) }{16B^{2}}\sin \vartheta \cos
\vartheta  \notag \\
& =\frac{J_{x}^{2}}{32B^{2}}\cos ^{2}\vartheta +\frac{\sqrt{\lambda
^{2}+\left( \lambda +J_{y}\right) ^{2}}}{32B^{2}}\sin ^{2}\left( \vartheta
-\vartheta _{0}^{y}\right)
\end{align}%
with%
\begin{equation}
\vartheta _{0}^{y}=-\arctan \left( \frac{\lambda +J_{y}}{\lambda }\right) .
\label{theta0}
\end{equation}%
It becomes zero for 
\begin{equation}
\vartheta =\vartheta _{0}^{y},\quad J_{x}=0,
\end{equation}%
where a perfect elliptic dichroism occurs. On the other hand, the perfect
elliptic dichroism does not occur for $J_{x}\neq 0$, but we will soon show
that the optical conductivity depends strongly on $\vartheta $.

\subsection{The case of\textbf{\ }$\mathbf{J}=(0,J_{y},0)$}

We study a specific case, where the N\'{e}el vector is along the $y$ axis.
The optical conductivity is exactly obtained as%
\begin{align}
& \frac{\sigma \left( \omega ;\vartheta \right) }{\sigma _{0}}  \notag \\
& =\frac{\pi \left\vert \lambda +J_{y}\right\vert \left( \hbar ^{2}\omega
^{2}+4B^{2}\right) }{2\lambda \hbar \omega ^{3}}\cos ^{2}\vartheta  \notag \\
& +\frac{\pi \lambda \left( \hbar ^{2}\omega ^{2}+4B^{2}\right) }{%
2\left\vert \lambda +J_{y}\right\vert \hbar \omega ^{3}}\sin ^{2}\vartheta -%
\frac{4B\pi }{\omega ^{2}}\cos \vartheta \sin \vartheta
\end{align}%
for $\hbar \omega >2\left\vert B\right\vert $, and $\sigma \left( \omega
;\vartheta \right) =0$ for $\hbar \omega <2\left\vert B\right\vert $. At the
optical band edge $\omega =2\left\vert B\right\vert $, we have%
\begin{align}
G(\mathbf{0};\vartheta )& =\frac{\left( \lambda +J_{y}\right) ^{2}}{32B^{2}}%
\cos ^{2}\theta +\frac{\lambda ^{2}}{32B^{2}}\sin ^{2}\theta  \notag \\
& -\frac{\lambda \left( \lambda +J_{y}\right) }{16B^{2}}\sin \theta \cos
\theta  \notag \\
& =\frac{\sqrt{\lambda ^{2}+\left( \lambda +J_{y}\right) ^{2}}}{32B^{2}}\sin
^{2}\left( \vartheta -\vartheta _{0}^{y}\right) ,
\end{align}%
and the optical conductivity is simplified as%
\begin{equation}
\frac{\sigma \left( 2\left\vert B\right\vert /\hbar ;\vartheta \right) }{%
\sigma _{0}}=\frac{\hbar ^{2}\sqrt{\lambda ^{2}+\left( \lambda +J_{y}\right)
^{2}}}{4B^{2}\left( \lambda +J_{y}\right) }\sin ^{2}\left( \vartheta
-\vartheta _{0}\right) .
\end{equation}%
It becomes zero at $\vartheta =\vartheta _{0}^{y}$ and $\vartheta
_{0}^{y}+\pi $, which leads to a perfect elliptic optical dichroism. By
searching the condition $\vartheta =\vartheta _{0}^{y}$, it is possible to
determine the ratio $\left( \lambda +J_{y}\right) /\lambda $ experimentally.
On the other hand, the optical conductivity takes the maximum value at $%
\vartheta =\vartheta _{0}^{y}\pm \pi /2$. The band gap $2\left\vert
B\right\vert $ is determined by the optical band edge. By combining it the
result of the maximum optical conductivity $\sigma \left( 2\left\vert
B\right\vert ;\vartheta _{0}^{y}\pm \pi /2\right) $, it is possible to
determine $B$, $\lambda $ and $J_{y}$ experimentally. In addition, it is
possible to differentiate two states $J_{y}$\ and $-J_{y}$\ by optical
absorption as shown in Fig.\ref{FigLinear}(a) and (b). This is one of the
main result of this paper. It is interesting that it is impossible to
differentiate $J_{x}$\ and $J_{z}$\ by linear polarized light with $\Theta
=0 $\ but is possible by that with $\Theta =\pi /2$,\ as shown in Fig.\ref%
{FigLinear}(a) and (b).

The optical conductivity at the optical band edge $\sigma \left( 2\left\vert
B\right\vert \right) $ is shown as a function of $\vartheta $ in Fig.\ref%
{FigEllipticPy}(a). There is a strong $\vartheta $ dependence. Especially,
it becomes exactly zero at $\vartheta =\vartheta _{0}^{y}$. The $\sigma
\left( \omega \right) $ is shown in Fig.\ref{FigEllipticPy}(b).

\begin{figure}[t]
\centerline{\includegraphics[width=0.49\textwidth]{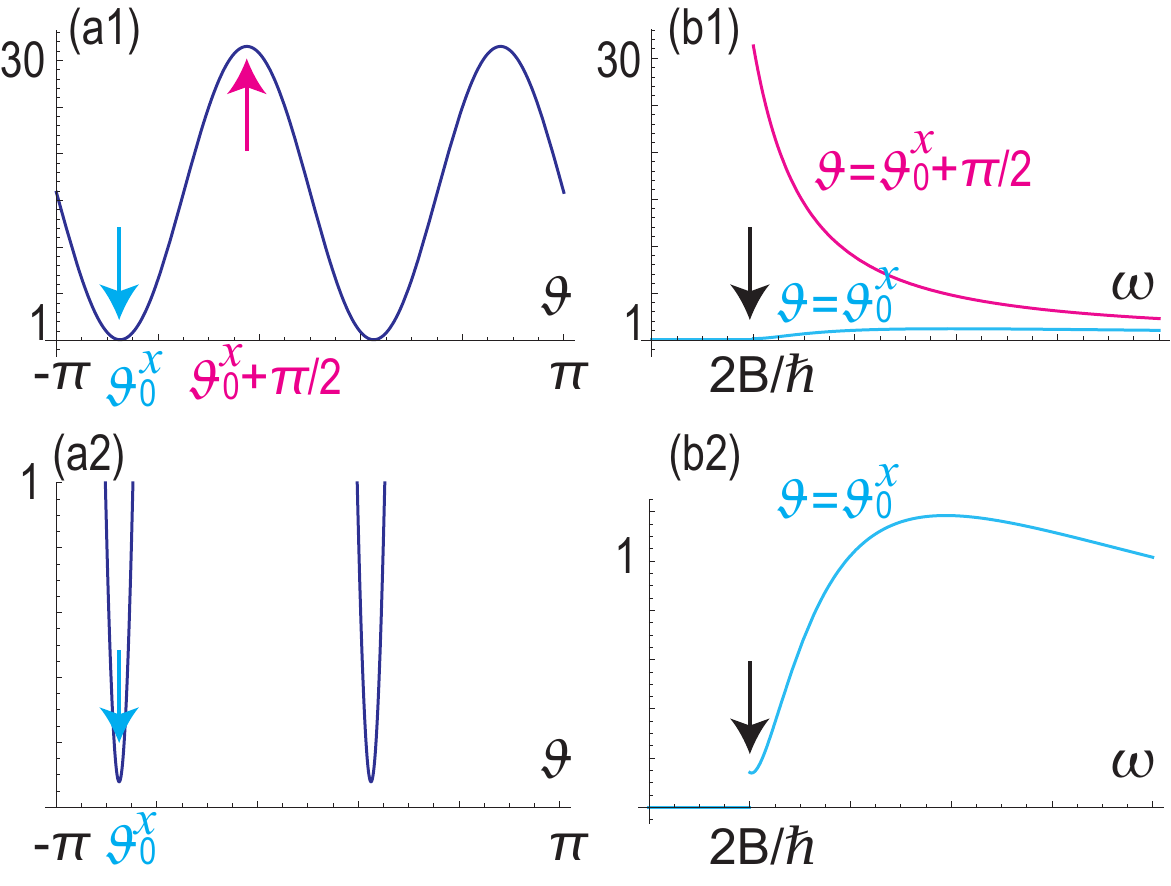}}
\caption{Elliptic dichroism, where the N\'{e}el vector is taken along the $x$
axis. (a2) and (b2) are the enlarged figures of (a1) and (b1) showing $%
\protect\sigma \left( 2\left\vert B\right\vert /\hbar ;\protect\vartheta %
_{0}^{y}\right) \neq 0$. See also the caption of Fig.3.}
\label{FigEllipticPx}
\end{figure}

\subsection{The case of\textbf{\ }$\mathbf{J}=(J_{x},0,0)$}

We study a specific case, where the N\'{e}el vector is along the $x$ axis.
The optical conductivity is exactly obtained as%
\begin{align}
& \frac{\sigma \left( \omega ;\vartheta \right) }{\sigma _{0}}  \notag \\
=& \frac{\pi \left( \lambda ^{2}+J_{x}^{2}\right) \left( \hbar ^{2}\omega
^{2}+4B^{2}\right) }{2\lambda ^{2}\hbar \omega ^{3}}\cos ^{2}\vartheta 
\notag \\
& +\frac{\pi \left( \hbar ^{2}\omega ^{2}+4B^{2}\right) }{2\hbar \omega ^{3}}%
\sin ^{2}\vartheta -\frac{4B\pi }{\omega ^{2}}\cos \vartheta \sin \vartheta
\end{align}%
for $\hbar \omega >2\left\vert B\right\vert $, and $\sigma \left( \omega
;\vartheta \right) =0$ for $\hbar \omega <2\left\vert B\right\vert $. It is
impossible to differentiate $J_{x}$\ and $-J_{x}$\ because the optical
absorption is quadratic in $J_{x}$\ as shown in Fig.\ref{FigLinear}(a) and
(b).

The optical conductivity at the optical band edge $\sigma \left( 2\left\vert
B\right\vert \right) $ is shown as a function of $\vartheta $ in Fig.\ref%
{FigEllipticPx}(a1). The optical conductivity at the optical band edge is
analytically obtained as%
\begin{equation}
\sigma \left( 2\left\vert B\right\vert /\hbar ;\vartheta \right) =\frac{%
\hbar ^{2}\pi J_{x}^{2}}{2B\lambda ^{2}}\cos ^{2}\vartheta +\frac{\hbar
^{2}\pi }{2B}\left( 1-\sin 2\vartheta \right) ,
\end{equation}%
where we have used the relation on the quantum geometry at the Dirac point,%
\begin{equation}
G(\mathbf{0};\vartheta )=\frac{J_{x}^{2}}{32B^{2}}\cos ^{2}\vartheta +\frac{%
\lambda ^{2}}{32B^{2}}\left( 1-\sin 2\vartheta \right) .
\end{equation}%
There is a strong dependence on $\vartheta $. It takes the minimum at 
\begin{equation}
\vartheta =\vartheta _{0}^{x}=-\arctan \left( \frac{2\lambda ^{2}}{J_{x}^{2}}%
\right) ,
\end{equation}%
where the minimum value is%
\begin{equation}
\sigma \left( 2\left\vert B\right\vert /\hbar ;\vartheta _{0}^{x}\right) =%
\frac{\pi \hbar ^{2}}{2B}\left( 2\lambda ^{2}+J_{x}^{2}+\sqrt{4\lambda
^{4}+J_{x}^{4}}\right) >0.
\end{equation}%
It does not become zero as shown in Fig.\ref{FigEllipticPx}(a2). The $\sigma
\left( \omega \right) $ is shown in Fig.\ref{FigEllipticPx}(b). The
behaviour is almost similar to that of the case where the N\'{e}el vector is
along the $y$ axis although there is a tiny nonzero contribution $\sigma
\left( \omega ;\vartheta _{0}+\pi /2\right) $\ at the optical band edge as
shown in Fig.\ref{FigEllipticPx}(b2).

\begin{figure}[t]
\centerline{\includegraphics[width=0.49\textwidth]{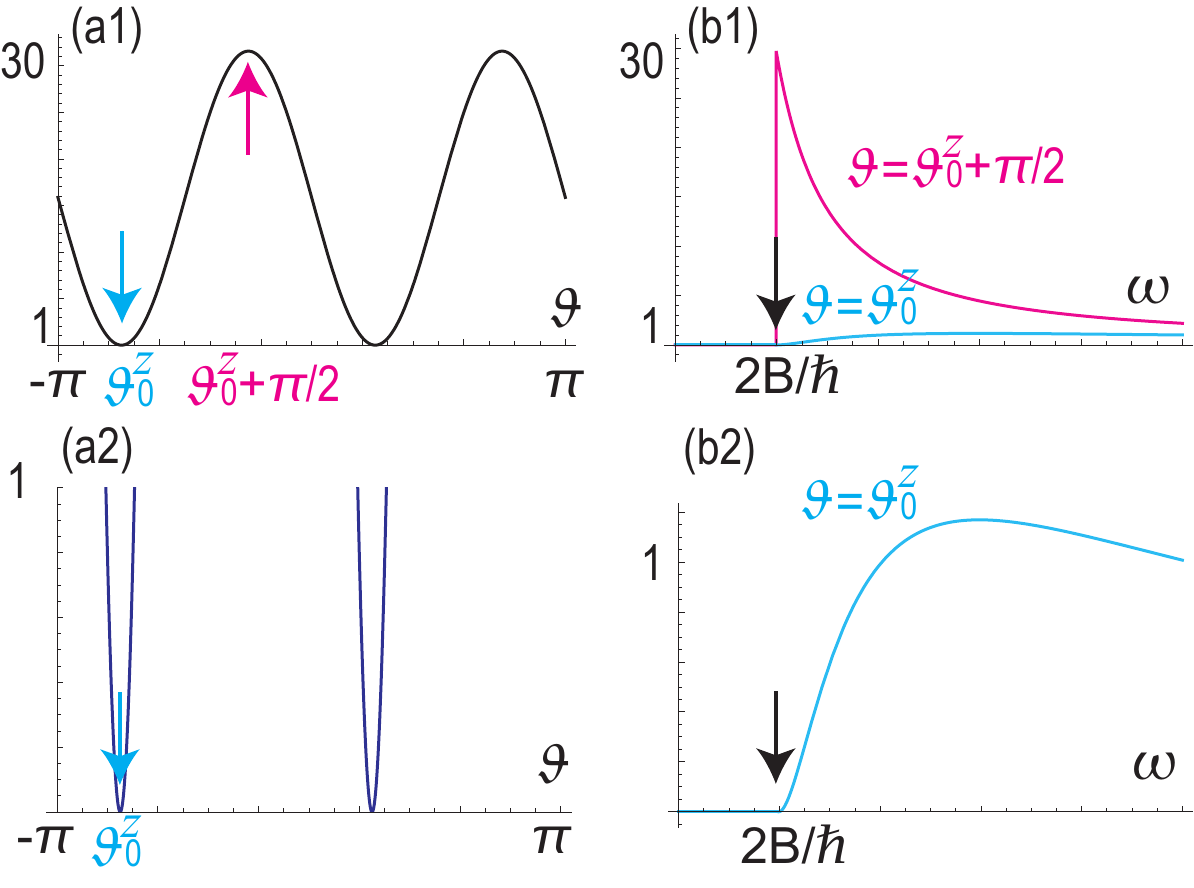}}
\caption{Elliptic dichroism, where the N\'{e}el vector is taken along the $z$
axis. See also the caption of Fig.3.}
\label{FigEllipticPz}
\end{figure}

\subsection{The case of\textbf{\ }$\mathbf{J}=(0,0,J_{z})$}

We study a specific case, where the N\'{e}el vector is along the $z$ axis.
The optical conductivity is exactly obtained as%
\begin{align}
\frac{\sigma \left( \omega ;\vartheta \right) }{\sigma _{0}}=& \Big[\frac{%
\pi \left( \hbar ^{2}\omega ^{2}+4B^{2}\right) }{2\hbar \omega ^{3}}+\frac{%
\pi J_{z}^{2}\left( \hbar ^{2}\omega ^{2}-4B^{2}\right) }{4\lambda ^{2}\hbar
\omega ^{3}}  \notag \\
& -\frac{\pi J_{z}^{4}\left( \hbar ^{2}\omega ^{2}-12B^{2}\right) }{%
16\lambda ^{4}\hbar \omega ^{3}}\Big]\cos ^{2}\vartheta  \notag \\
& +\Big[\frac{\pi \left( \hbar ^{2}\omega ^{2}+4B^{2}\right) }{2\hbar \omega
^{3}}-\frac{\pi J_{z}^{2}\left( \hbar ^{2}\omega ^{2}+12B^{2}\right) }{%
4\lambda ^{2}\hbar \omega ^{3}}  \notag \\
& +\frac{3\pi J_{z}^{4}\left( \hbar ^{2}\omega ^{2}+20B^{2}\right) }{%
16\lambda ^{4}\hbar \omega ^{3}}\Big]\sin ^{2}\vartheta  \notag \\
& -\frac{B\pi }{\omega ^{2}}\Big(4-\frac{2J_{z}^{2}}{\lambda ^{2}}+\frac{%
3J_{z}^{4}}{2\lambda ^{4}}\Big)\cos \vartheta \sin \vartheta
\end{align}%
for $\hbar \omega >2\left\vert B\lambda \right\vert /\sqrt{\lambda
^{2}+J_{z}^{2}}$, and $\sigma \left( \omega ;\vartheta \right) =0$ for $%
\hbar \omega <2\left\vert B\lambda \right\vert /\sqrt{\lambda ^{2}+J_{z}^{2}}
$. It is impossible to differentiate $J_{z}$\ and $-J_{z}$\ because the
optical absorption is a function of $J_{x}^{2}$\ as shown in Fig.\ref%
{FigLinear}(a) and (b).

The optical conductivity at the optical band edge $\sigma \left( 2\left\vert
B\right\vert \right) $ is shown as a function of $\vartheta $ in Fig.\ref%
{FigEllipticPz}(a). There is a strong $\vartheta $ dependence. Especially,
it becomes exactly zero at 
\begin{equation}
\vartheta \simeq \vartheta _{0}^{z}=-\arctan \left( \frac{\lambda +J_{z}}{%
\lambda }\right) .
\end{equation}%
The $\sigma \left( \omega \right) $ is shown in Fig.\ref{FigEllipticPz}(b).

\section{Discussions}

The $p$-wave magnets have the zero net magnetization. It is expected to be
useful for antiferromagnetic spintronics\cite%
{Jung,Baltz,Han,Ni,Godin,Kimura,ZhangNeel} with high density and fast
switchable magnetic memories because the effect of the stray field is small
comparing with ferromagnets. In this context, it is important to
differentiate the up and down states of the N\'{e}el vector. It was shown%
\cite{EzawaPNeel} that the N\'{e}el vector is detectable by measuring the
linear and nonlinear conductivities. We discuss a merit of optical detection
of the N\'{e}el vector comparing conductivities. It is possible to observe
spatial dependent optical absorption by a single shot experiment, where it
is possible to observe the domain wall structure of the $p$-wave magnet. It
is interesting to observe a domain-wall motion of the $p$-wave magnet by
optical absorption.

By using the perfect elliptic dichroism condition (\ref{theta0}), it is
possible to determine $J_{y}$ if $\lambda $ is known. Hence, it is possible
to differentiate the two states $J_{y}$ and $-J_{y}$ by optical measurement.
On the other hand, it is impossible to differentiate $J_{x}$\ and $-J_{x}$\
because the optical absorption is quadratic in $J_{x}$. This is also the
case for $J_{z}$. In addition, it is hard to determine the direction of the N%
\'{e}el vector only by elliptic dichroism because the results are similar as
shown in Figs.\ref{FigEllipticPy},\ref{FigEllipticPx} and \ref{FigEllipticPz}%
. In particular, the optical conductivity takes the minimum value at $%
\vartheta =\vartheta _{0}$ irrespective of the direction of the N\'{e}el
vector. Nevertheless, our results will be useful for future spintronics
memories based on the $p$-wave magnet because the direction of the N\'{e}el
vector is fixed by the shape anisotropy of the sample and there are only two
stable directions.

The author is very much grateful to M. Hirschberger and S. Okumura for
helpful discussions on the subject. This work is supported by CREST, JST
(Grants No. JPMJCR20T2) and Grants-in-Aid for Scientific Research from MEXT
KAKENHI (Grant No. 23H00171).

\end{document}